\documentclass[12pt]{JHEP}
\usepackage{graphics}
\usepackage{cite}
\newcommand{\beq}{\begin{equation}}
\newcommand{\eeq}{\end{equation}}
\newcommand{\bea}{\begin{eqnarray}}
\newcommand{\eea}{\end{eqnarray}}

\renewcommand{\d}{\delta}
\renewcommand{\l}{\lambda}
\renewcommand{\L}{\Lambda}
\renewcommand{\b}{\beta}

\newcommand{\m}{\mu}

\newcommand{\s}{\sigma}
\newcommand{\D}{{\cal D}}
\renewcommand{\S}{{\cal S}}

\newcommand{\tf}{\tilde{f}}
\newcommand{\V}{{\cal V}}

\newcommand{\oh}{\frac{1}{2}}

\newcommand{\dg}{\dagger}
\newcommand{\non}{\nonumber}

\newcommand{\rf}[1]{(\ref{#1})}
\newcommand{\ra}{\rightarrow}

\title{Direct Laplacian Center Gauge}

\author{Manfried Faber \\ Atominstitut der \"osterreichischen
Universit\"aten, \\
Arbeitsgruppe Kernphysik, TU Wien, 
A-1040 Vienna, Austria. \\
E-mail: \email{faber@kph.tuwien.ac.at}}

\author{Jeff Greensite \\ The Niels Bohr Institute, Blegdamsvej 17,
DK-2100 Copenhagen \O, Denmark. \\
E-mail: \email{greensit@alf.nbi.dk} \\
Physics and Astronomy Department, San Francisco
State University, San Francisco, CA 94117 USA. 
E-mail: \email{greensit@quark.sfsu.edu}}

\author{{\v S}tefan Olejn\'{\i}k \\ Institute of Physics, Slovak Academy 
of Sciences, SK-842 28 Bratislava, Slovakia. 
E-mail: \email{fyziolej@savba.sk}}

\abstract{We introduce a variation of direct maximal center gauge
fixing: the ``direct Laplacian'' center gauge.  The new procedure 
overcomes certain shortcomings of maximal center gauge,
associated with Gribov copies, that were pointed out 
by Bornyakov et al.\ in hep-lat/0009035.}


\preprint

\begin{document}

\section{Introduction}

   Center projection in direct maximal center gauge is a procedure
used to locate center vortices in lattice gauge field configurations.
Results obtained from this procedure, which tend to support the center vortex
theory of confinement, have been reported in many articles over the past
several years.  Last
year, however, an alarming negative result was reported by Bornyakov,
Komarov, and Polikarpov (BKP) in ref.\ \cite{BKP}.  Recall that
direct maximal center gauge fixing is defined by the gauge 
transformation $g$ which maximizes the sum 
\beq
      R = \sum_{x,\m} \mbox{Tr}_A[{}^gU_\m(x)]
\label{R}
\eeq
where $\mbox{Tr}_A$ indicates the trace in the adjoint representation,
and ${}^gU_\m(x)$ is the gauge-transformed lattice configuration.
This prescription is simply Landau gauge fixing in the adjoint
representation.  There is no known method for finding the global
maximum of $R$, but most previous work has used the over-relaxation
scheme of ref.\ \cite{Jan98} to find local maxima (Gribov copies).
The string tension that can be attributed to vortices, which are
located using this scheme, agrees quite well with the asymptotic
string tension of the full lattice configuration.  This agreement is
known as ``center dominance,'' and is crucial to the argument that
center vortices account for the entire asymptotic string tension.  In
their work last year, BKP used an improved center gauge-fixing scheme
based on simulated annealing, and obtained Gribov copies with
consistently larger values of $R$ than those obtained by
over-relaxation method.  The discouraging outcome of this improved
gauge-fixing procedure is that center dominance is much less evident:
the center projected SU(2) string tension at $\b=2.5$ is some $30\%$
lower than than the full asymptotic string tension.

   In this article we explain the origin of the difficulty found by
BKP, and then propose and test a method which overcomes that
difficulty.  The 
realization that direct maximal center
gauge must fail in the continuum limit is actually due to Engelhardt and
Reinhardt in ref.\ \cite{ER}, whose work predates the numerical
results of BKP in ref.\ \cite{BKP}.  We have previously elaborated on the
Engelhardt-Reinhardt reasoning
in ref.\ \cite{Remarks}; the salient points will be reviewed
in section 2 below.  In brief, these authors pointed out
that direct maximal center gauge can be understood as a best fit to a
given lattice gauge field by a thin vortex configuration.  Since the
field strength of a thin vortex is highly singular, the fit must fail
badly near the middle of the vortex.  In the continuum limit, the bad
fit to the gauge field in the vortex interior would overwhelm the good
fit in the exterior region.  As a result, in the continuum limit, the
center projection obtained from a global maximum of $R$ may reveal
no vortices at all.

   Having diagnosed the illness, a cure is proposed in section 3.
We work throughout with the SU(2) gauge group,
although the analysis generalizes readily to SU(N).
The new method calls for maximizing
\beq
      R_M = \sum_{x,\m} \mbox{Tr}[M^T(x) U_{A\m}(x) M(x+\hat{\m})]
\eeq
where $U_{A\m}(x)$ denotes the link variable in the adjoint representation,
and where, for SU(2) gauge theory, $M(x)$ is a real-valued $3\times 3$ 
matrix which satisfies the unitarity constraint only as an average; i.e.
\beq
        {1\over \V} \sum_x \sum _j M^T_{ij}(x) M_{jk}(x) = \d_{ik}
\eeq
with $\V$ the lattice volume.  The motivation for relaxing the
local unitarity constraint is explained below.  The real-valued 
matrix field $M(x)$ which
arrives at a global maximum of $R_M$ can be determined uniquely, in
terms of the three lowest eigenfunctions of a lattice Laplacian
operator.  To determine the corresponding gauge transformation, we
relax $M(x)$ to the closest SO(3) matrix-valued field $g_A(x)$
satisfying a related Laplacian condition.  The SO(3) field is mapped
to an SU(2) matrix-valued field $g(x)$, which is used to gauge-transform
the original lattice configuration.  Center projection then 
determines the vortex locations.  

   In the end, direct Laplacian center gauge is simply lattice Landau gauge,
in the adjoint representation, with a particular choice of Gribov copy
(different from the global maximum of $R$).  We will motivate this
choice in sections 2 and 3.  Numerical results obtained from the new
procedure, regarding center
dominance, vortex density scaling, precocious linearity, and the
correlation of vortex locations with the values of Wilson loops, are
reported in section 4.

   Direct Laplacian center gauge is closely related to the original
Laplacian Landau gauge of Vink and Wiese \cite{Vink}, appropriately
generalized to the adjoint representation.  The main difference is in
the mapping from $M(x)$ to $g(x)$.  Direct Laplacian gauge also
has much in common with the Laplacian center gauge, introduced by
Alexandrou et al.\ \cite{Alex} and de Forcrand and Pepe \cite{Pepe}, 
in that both gauges
require solving the lattice Laplacian eigenvalue problem  in the
adjoint representation.  But the gauges themselves are rather different; 
they are motivated by different considerations (best fit in one case,
gauge-fixing ambiguity in the other), and the numerical results are
not the same.  These points are further discussed
in section 5.  Section 6 contains our conclusions.

\section{Maximal Center Gauge and its Discontents}

   We begin from the insight that lattice Landau gauge fixing is
precisely equivalent to finding the best fit, to a given lattice
gauge field $U_\m(x)$, by a classical vacuum configuration 
$g(x)g^\dg(x+\hat{\m})$.  The best fit vacuum configuration
minimizes the mean square
distance on the group manifold between vacuum and gauge field link variables
\bea
     d^2_F &=& {1\over 4\V} \sum_{x,\m} \mbox{Tr} \left[
     \left( U_\m(x) - g(x)g^\dg(x+\hat{\m}) \right) 
     \left( U_\m(x) - g(x)g^\dg(x+\hat{\m}) \right)^\dg \right]
\non \\
           &=& {1\over 4\V} \sum_{x,\m} 2 \mbox{Tr} \left[
                I - g^\dg(x) U_\m(x) g(x+\hat{\m}) \right]
\label{d2F}
\eea
Writing
\beq
       {}^gU_\m(x) = g^\dg(x) U_\m(x) g(x+\hat{\m})
\eeq
it is is clear that minimizing \rf{d2F} is equivalent to maximizing
\beq
         \sum_{x,\m} \mbox{Tr}[{}^gU_\m(x)]
\eeq
in the fundamental representation, which is just the lattice Landau gauge.

   Generalizing the above idea slightly, suppose we are interested in
finding the best fit of $U_\m(x)$ by a thin center vortex configuration,
which has the form
\beq
       V_\m(x) = g(x) Z_\m(x) g^\dg(x+\hat{\m})
\label{V}
\eeq
where the $Z_\m(x)=\pm 1$ are $Z_2$ link variables.  Since the adjoint
representation is blind to center elements, we may proceed in the
following way:  First find the vacuum configuration
$g(x) g^\dg(x+\hat{\m})$ which is a best fit to $U_\m(x)$ in the
adjoint representation, i.e.\ which minimizes
\bea
     d^2_A &=& {1\over 4\V} \sum_{x,\m} \mbox{Tr}_A \left[
     \left( U_\m(x) - g(x)g^\dg(x+\hat{\m}) \right) 
     \left( U_\m(x) - g(x)g^\dg(x+\hat{\m}) \right)^\dg \right]
\non \\
           &=& {1\over 4\V} \sum_{x,\m} 2 \mbox{Tr}_A \left[
                I - g^\dg(x) U_\m(x) g(x+\hat{\m}) \right]
\label{d2A}
\eea
It is clear that the minimum of $d^2_A$ is obtained at the maximum
of
\beq
          R = \sum_{x,\m} \mbox{Tr}_A[{}^gU_\m(x)]
\eeq
which is, by definition, direct maximal center gauge.  Having determined
$g(x)$ in this way (up to a residual $Z_2$ symmetry), we then
want to find the $Z_\m(x)$ such that the thin vortex configuration
$V_\m(x)$ is a best fit to $U_\m(x)$.  This means minimizing, in
the fundamental representation
\bea
     d^2 &=& {1\over 4\V} \sum_{x,\m} \mbox{Tr} \left[
     \left( U_\m(x) - g(x)Z_\m(x)g^\dg(x+\hat{\m}) \right) 
     \left( U_\m(x) - g(x)Z_\m(x)g^\dg(x+\hat{\m}) \right)^\dg \right]
\non \\
           &=& {1\over 4\V} \sum_{x,\m} 2 \mbox{Tr} \left[
                I - Z_\m(x) g^\dg(x) U_\m(x) g(x+\hat{\m}) \right]
\eea
For fixed $g(x)$, minimization is achieved by setting
\beq
     Z_\m(x) = \mbox{signTr}[{}^gU_\m(x)]
\label{Z}
\eeq
which is the center projection prescription.  
In this way we see that maximal
center gauge, plus center projection, is equivalent to finding
a ``best fit'' of the given lattice gauge field $U_\m(x)$ by a
thin vortex configuration $V_\m(x)$.  This point was first made by
Engelhardt and Reinhardt \cite{ER} in the context of continuum
Yang-Mills theory; here we have transcribed their argument to the
lattice.

   Unfortunately, it is clear that $V_\m(x)$ must be a very bad
fit to $U_\m(x)$ at links belonging to thin vortices (i.e.\ to the
P-plaquettes formed from $Z_\m(x)$).  Again, this point was made
in ref.\ \cite{ER} in the context of the continuum theory, on the
grounds that the field strength of a thin vortex is divergent at the
vortex core.  We recall that a 
plaquette $p$ is a P-plaquette iff $Z(p)=-1$ (where $Z(C)$ denotes the
product of $Z_\m(x)$ around the contour $C$) and that  P-plaquettes belong
to P-vortices.  This means also that 
\beq
       \oh \mbox{Tr}[V(p)] = Z(p) = -1
\label{pplaq}
\eeq
On the other hand, at large $\b$ we generally have
\beq
      \oh \mbox{Tr}[U(p)] =  1 - O\left({1\over \b}\right)
\label{largebeta}
\eeq
even at P-plaquettes.  Writing link variables $U_\m(x)$ in the
form
\bea
       U_\m(x) &=& g(x)Z_\m(x) e^{iA_\m(x)} g^\dg(x+\hat{\m})
\non \\
   {}^gU_\m(x) &=& Z_\m(x) e^{iA_\m(x)} 
\eea
we have
\beq
 \mbox{Tr}[U(p)] = Z(p) \mbox{Tr}[\prod_{links \in p}e^{iA_\m(x)}]
\eeq
or, in view of \rf{pplaq},\rf{largebeta}, 
\beq
      \oh \mbox{Tr}[\prod_{links \in p}e^{iA_\m(x)}] \approx -1
\eeq
But, from eq.\ \rf{Z},
\beq
       \mbox{Tr}\left[ e^{iA_\m(x)}\right] \ge 0
\eeq
on every link. Taken together, the last two equations imply that for some links
(at least one) on a P-plaquette, ${}^gU_\m(x)$ deviates strongly away from the
center elements $\pm I$, and, as a result,
\beq
      \mbox{Tr}_A\left[{}^gU_\m(x) \right] \ll
          \mbox{Tr}_A\left[I\right]
\eeq
It follows that the trace of links in the adjoint representation,
which is the quantity being maximized in direct maximal center gauge, 
will be much smaller, on average, in the vicinity of a P-plaquette than
on the rest of the lattice.

   If we compare the fits to a thermalized lattice that can be obtained
from a vacuum configuration (no P-vortices), and from 
a configuration containing some arrangement of thin vortices on 
vacuum background, the latter is likely to be a better fit in the lattice
region exterior to P-plaquettes, but a much worse fit at the P-plaquettes
themselves.  In the $\b\ra \infty$ limit, the bad fit near the P-plaquettes
may overwhelm the good fit in the exterior region, particularly if
the thermalized lattice contains thick ($\approx$ 1 fm) 
center vortices which overlap
substantially.  In that case, the best fit is just a pure-gauge, and the
global maximum of $R$ would reveal no vortices at all.

   In our opinion, this is the explanation for the discouraging result
found by BKP in ref.\ \cite{BKP}.  Using the simulated annealing method
described in that reference, we have found that center projection in
direct maximal center gauge has good center dominance properties at
strong couplings, but that center dominance degrades as we enter the
scaling regime.  We have also found that Gribov copies obtained by
our original method of over-relaxation arrive at a better fit to the
thermalized lattice, as compared to copies generated by simulated
annealing, if the computation of $R$ is restricted to plaquettes in
the region exterior to P-plaquettes.  This is consistent with the
idea that the bad fit in the neighborhood of P-plaquettes is the source of
the trouble. For details and numerical results 
regarding this point, we refer the reader to ref.\ \cite{Remarks}.

   Having arrived at a diagnosis of the problem, the question is what 
to do about it.  One possibility is simply to continue using the original
over-relaxation technique, on the grounds that it provides a better fit
to the lattice in the region exterior to P-plaquettes.  Unfortunately,
if one follows this idea further, gauge-fixing a large number of random
copies on the gauge orbit and selecting the copy with the best fit in 
the exterior region, the string tension comes out too high \cite{Remarks}.
Therefore, we cannot recommend the ``business-as-usual'' approach with
much enthusiasm.

   Another possibility, suggested by Engelhardt and Reinhardt \cite{ER},
is to modify $R$ in eq.\ \rf{R} by introducing a form factor $F$ into 
the expression, i.e.
\beq
      R^F = \sum_{x,\m} F\Bigl[ \mbox{Tr}_A[{}^gU_\m(x)] \Bigr]
\label{RF}
\eeq
In particular, one could try the ``lower bound'' function
\beq
        F[x] = \left\{ \begin{array}{cl}
                           x & ~~~ x > \Lambda \cr
              \Lambda      & ~~~ x \le \Lambda 
              \end{array} \right.
\label{F}
\eeq
in an effort to soften the bad fit at P-plaquettes.
We have experimented with this form of $F$, but the whole approach is
obviously plagued by arbitrariness.  With an extra
free parameter such as $\Lambda$ (that can be reset at each $\b$),
it is not surprising that one can 
adjust the projected string tension to the desired result.  

   Finally, there is the Laplacian center gauge \cite{Alex} developed by de
Forcrand and Pepe \cite{Pepe}, which is free of Gribov ambiguities
altogether.  Our present article is inspired in large part by the de
Forcrand-Pepe approach (as well as by the earlier work of Vink and
Wiese \cite{Vink}), but we do have some reservations about the
version of Laplacian center gauge discussed in refs.\ \cite{Alex,Pepe}.
These reservations concern the lack of scaling of the vortex density,
the lack of precocious linearity in the vortex potential, and the
prescription for locating vortices from the co-linearity of two
Laplacian eigenvectors, which appears to lack the important
``vortex-finding property'' for thin vortices \cite{vf}.  This last
point, which is relevant to the program of looking for gauge-fixing
ambiguities in Laplacian gauges, will be discussed in section 5.  In
the meantime, we will proceed to the proposal which is main point of
this article.
           
\section{Direct Laplacian Center Gauge}

   To motivate our proposal, we return temporarily 
to the idea that, since the fit is worst at P-plaquettes,
it might be sensible to exclude those contributions from the
quality-of-fit function $R$.  Specifically,
consider introducing a configuration-dependent weighting factor 
\beq
       \rho(x;{}^gU_\m) = \left\{ \begin{array}{cl}
    0 & ~~~ x \in \mbox{P-vortex} \cr
    c & ~~~ \mbox{otherwise} \end{array} \right.
\eeq
where center projection of ${}^gU_\m$ is used to decide whether or not
the site $x$ belongs to a P-plaquette, and $c$ is a constant  
determined by the constraint
\beq
    {1\over \V} \sum_x \rho^2(x;{}^gU_\m) = 1
\eeq
which means that $c^2$ is inversely proportional to the lattice volume
exterior to P-plaquettes.
One then considers
\beq
      R' =  \sum_{x,\m} \rho(x;{}^gU_\m) \rho(x+\hat{\m};{}^gU_\m)
         \mbox{Tr}_A[g^\dg(x) U_\m(x) g(x+\hat{\m})]
\label{Rrho}
\eeq
which is proportional to the average trace of adjoint links
$\mbox{Tr}_A[{}^gU_\m(x)]$ in the exterior volume,
choosing $g(x)$ to maximize this quantity.

   However, even the exclusion of all links joined to P-plaquettes is
still not good enough in the continuum $\b \ra \infty$ limit, since
there one expects a bad fit also in a finite volume surrounding the
P-plaquettes. At low or intermediate values of $\b$, on the other
hand, excluding 100\% of the P-plaquette contributions goes too far,
because the excluded links are a very substantial fraction of the
total lattice volume.  The next step, then, is to allow greater
flexibility in the weighting factor.  This is done by allowing
$\rho(x)$ to be a degree of freedom in its own right.  Introduce a
$3\times 3$ matrix-valued field
\beq
      G_{ij}(x) = \pm \Omega_{ij}(x) \rho(x)
\eeq
where $\Omega$ is an SO(3) matrix, and consider choosing $\rho,\Omega$ to
maximize the quantity
\beq
      R'' = \sum_{x,\m} \mbox{Tr}[G^T(x) U_{A\m}(x) 
         G(x+\hat{\m})] 
\label{R'}
\eeq
where $U_{A\m}(x)$ denotes the link variable in the adjoint representation,
and where $\rho(x)$ is
a real positive scalar, subject to the constraint
\beq
      {1 \over \V} \sum_x \rho^2(x) = 1
\label{constraint}
\eeq
This constraint is now simply the requirement that $G(x)$ is orthogonal
``on average.''  
Mapping the SO(3) matrix field $\Omega(x)$ onto an $SU(2)$ matrix
field $g(x)$ with, e.g., Tr$[g]>0$, 
and again defining the lattice field in this new gauge 
by ${}^gU_\m(x)$, one can determine the P-vortex locations from the center 
projected configuration \rf{Z}.

   Generalizing this stategy one step further, we can replace the
positive scalar field $\rho(x)$
by a real, symmetric matrix $P_{ij}(x)$ with positive semidefinite
eigenvalues, and write
\beq
          M_{ij}(x) = \pm \Omega_{ik}(x)  P_{kj}(x)
\eeq
(summation is over repeated indices).
This is known as the polar decomposition of the matrix $M$.  A corollary
of the Singular Value Decomposition Theorem \cite{Golub} is that
any real $3 \times 3$ matrix $M$ can be decomposed in this way.
The idea is then to find the $3 \times 3$ matrix-valued field  $M(x)$
which maximizes 
\beq
       R_M = \sum_{x,\m} \mbox{Tr}[M^T(x) U_{A\m}(x) 
         M(x+\hat{\m})] 
\label{Rt}
\eeq
subject to the constraint (which generalizes \rf{constraint})
\beq
     {1\over \V} \sum_x M^T(x) M(x) = I
\label{constraint1}
\eeq
so that $M(x)$ is also orthogonal ``on average.''

   It is fortunate that the problem we have just posed has both a unique
solution, and a standard computational algorithm for arriving at
that solution.  It is convenient to view the columns of $M(x)$
at any site $x$ as a set of three 3-vectors
\beq
       f^b_a(x) = M_{ab}(x) ~~,~~ \vec{f^b}(x) \equiv \left[ \begin{array}{c}
          f_1^b(x) \cr f_2^b(x) \cr f_3^b(x) \end{array} \right]
\eeq
Then in this notation, maximizing $R_M$ is equivalent to maximizing
\beq
  S = \sum_{x,\m} f^c_j(x) [U_{A\m}(x)]_{jk} f^c_k(x+\hat{\m}) 
          + \oh \L_{cd} \sum_x \left[f^c_j(x) f^d_j(x) - \d_{cd} \right]
\label{S}
\eeq
where the $\L$ is a real symmetric matrix of Lagrange multipliers, 
introduced to enforce the constraint
\rf{constraint1}.  In view of the constraint, the configuration maximizing
$S$ also maximizes the expression
\beq
      S' = \sum_{x,\m} \left[f^c_j(x) [U_{A\m}(x)]_{jk} f^c_k(x+\hat{\m})
 - f^c_j(x) f^c_j(x) \right]
         + \oh \L_{cd} \sum_x \left[f^c_j(x) f^d_j(x) - \d_{cd} \right]
\label{S'}
\eeq
Variation of $S'$ wrt $f_i^a$ then leads to a lattice Laplacian
equation
\beq
      \sum_y \D_{ij}(x,y) f^a_{j}(y) = \L_{ac} f^c_i(x)
\label{laplace}
\eeq
where
\beq
  \D_{ij}(x,y)  = - \sum_{\mu}\left( 
         [U_{A\m}(x)]_{ij}\delta_{y,x+\hat\mu}
       + [U_{A\m}(x-\hat\mu)]_{ji}\delta_{y,x-\hat\mu}
      -  2 \delta_{ij} \delta_{xy}\right).
\label{D}
\eeq
Because $\L$ is real and symmetric, it can be diagonalized by an
orthogonal matrix $O$, i.e.
\bea
       \L &=& O^T \L_D O ~~, ~~~~ \L_D = \mbox{diag}[\l_1,\l_2,\l_3]
\non \\
       f'^a_i(x) &=& O_{ab} f^b_a(x)
\eea
and then the equation satisfied by each $\vec{f'^a}(x)$ 
is the Laplacian eigenvalue equation
\beq
      \sum_y \D_{ij}(x,y) f'^a_{j}(y) = \l_a f'^a_i(x) 
 ~~~~ \mbox{(no sum over $a$)}
\label{eigen}
\eeq

  Since $\D_{ij}(x,y)$ is hermitian, the orthogonality constraint
\rf{constraint1} is satisfied by choosing the $\l_a$ to be three
different eigenvalues.  Substituting the solutions of \rf{eigen} back
into \rf{S'}, its not hard to see that $S'$ is maximized by choosing
eigenvectors corresponding to the three \emph{lowest} eigenvalues.  An
efficient numerical algorithm for obtaining the low-lying eigenvectors
of a large, sparse matrix (such as our lattice Laplacian) is the
Arnoldi method \cite{Golub}. Conveniently, Fortran routines
implementing this method, and easily adaptable to the problem at hand,
are freely available \cite{ARPACK}.

   Since $S'$ is invariant under the global transformation 
$f^a_i(x) \ra f'^a_i(x) = O_{ab} f^b_a(x)$,
we may as well use the solution for the maximum for which
$M_{ab}(x)=f'^b_a(x)$.
At this point, we have to map the matrix field $M(x)$ onto an
SO(3)-valued field $g_A(x)$, and this in turn to a corresponding SU(2)
gauge transformation $g(x)$.  
We consider two possibilities:
\begin{enumerate}
\item {\bf Naive Map:~} Choose $g_A(x)$ to be the SO(3)-valued field
which is closest to $M(x)$, in the sense that
\beq
      \Bigl| \mbox{Tr}[g_A(x) M^T(x)] \Bigr| 
\label{close}
\eeq
is maximized at each site $x$.
\item {\bf Laplacian Map:~} Choose $[g_A(x)]_{ij} = \tilde{f}_i^j(x)$
to be the SO(3)-valued field closest to $M(x)$ subject to the
constraint that $g_A(x)$, like $M(x)$, satisfies a Laplacian equation
of the form
\beq
      \sum_y \D_{ij}(x,y) \tf^a_{j}(y) = \L_{ac}(x) \tf^c_i(x)
\label{glaplace}
\eeq
\end{enumerate}

  The ``naive'' choice is equivalent to the Laplacian Landau gauge
of Vink and Wiese \cite{Vink}, generalized to the adjoint representation.
It is obtained by first making a singular value decomposition
\beq
       M(x) = U(x) M_D(x) V^T(x)
\eeq
with $U,V$ orthogonal matrices 
and $M_D$ a diagonal positive semi-definite matrix.\footnote{Again, there
are standard numerical packages which implement singular value decomposition.
We have used the routine {\tt svdcmp} from Numerical Recipes \cite{Press}.}
Then make the polar decomposition at each site
\beq
      M(x) = [U(x) V^T(x)] [V(x) M_D(x) V^T(x)] \equiv \pm g_A(x) P(x)
\eeq
with $g_A(x)$ the SO(3) matrix-valued field
\beq
   g_A(x) = \det[U(x) V^T(x)] U(x) V^T(x)
\eeq
and $P$ is symmetric and positive semi-definite.
Matrix $g_A(x)$ is guaranteed to be the SO(3) matrix 
closest to $M(x)$, in the sense of maximizing the expression \rf{close}
\cite{Golub}. Finally, we map $g_A(x)$ at each site onto one of its two 
possible representatives in the SU(2) group
\beq
      g(x) = \pm {I + \s^a \s^b [g_A(x)]_{ab} \over 
                2[1 + \mbox{Tr}(g_A(x))] }
\label{map}
\eeq
The choice of sign in \rf{map} is irrelevant, all choices being related
by the residual $Z_2$ 
invariance.  One can then transform the lattice configuration by
$g(x)$ to obtain ${}^gU_\m(x)$, and carry out center projection.

   The drawback of the ``naive'' choice is that, while $M(x)$ is
a covariantly smooth matrix, whose columns are the low-lying eigenvectors
of the lattice Laplacian, this covariant smoothness property is
less pronounced in the $g_A(x)$ obtained from singular value
decomposition.  This results in high-frequency ``noise'' (small
scale fluctuations) in the P-vortex surfaces of the projected
configuration.  We have found the Laplacian mapping to be preferable.

   To carry out the Laplacian mapping, we begin with the ``naive''
mapping, to find the SO(3) matrix field $\Omega(x)$ closest to
the real-valued matrix field $M(x)$ obtained from solving the
Laplacian eigenvalue equation \rf{eigen}.  Then we relax $\Omega(x)$
to the closest solution $g_A(x)$ satisfying the generalized Laplacian
equation \rf{glaplace}.  The relaxation is carried out in the following
way:  We note that the equation \rf{glaplace} is simply the stationarity
condition of the action
\beq
      S'' = \sum_{x,\m} \left[f^c_j(x) [U_{A\m}(x)]_{jk} f^c_k(x+\hat{\m})
 - f^c_j(x) f^c_j(x) \right]  
    + \oh \sum_x \L_{cd}(x) \left\{f^c_j(x) f^d_j(x) - \d_{cd} \right\} 
\label{S''}
\eeq
with $[g_A(x)]_{ij} = f_i^j(x)$ an SO(3) field.  With the local
Lagrange multiplier field enforcing the SO(3) constraint, a stationary
solution of $S''$ is easily seen to be a Gribov copy of direct
maximal center gauge; local maxima of $S''$ are also
local maxima of $R$ in eq.\ \rf{R}.

   If $\S[g] \equiv -R$ is regarded as the action of a spin system, with
$g_A(x)$ the SO(3)-valued field variables, 
then each Gribov copy can be regarded
as a metastable state of system, each with its own ``basin of attraction.''
The basin of attraction of a metastable state is the volume of
all configurations which, when the system is suddenly cooled (or ``quenched''),
will fall into that metastable state.  Over-relaxation is essentially
a sudden cooling technique \cite{Remarks}, and is therefore perfectly suited
to sliding the system ``down the hill'' of action $\S$, from the
configuration ${}^\Omega U_{A\m}(x)$ to
the nearest (or, at least, a nearby) local minimum of $\S$ 
(see Fig. 1).\footnote{There is
no guarantee that the local minimum obtained in this way is
truly the nearest of the local minima to $M(x)$, as this depends on the
topography of $\S[g]$ in the neighborhood of $\Omega$.}
The gauge transformation $g(x)$ obtained at this minimum 
is the Laplacian map of $M(x)$, and the corresponding ${}^g U_\m(x)$ 
is the lattice configuration in direct Laplacian center gauge.
   
\FIGURE[h!]{
\centerline{\scalebox{0.75}{\includegraphics{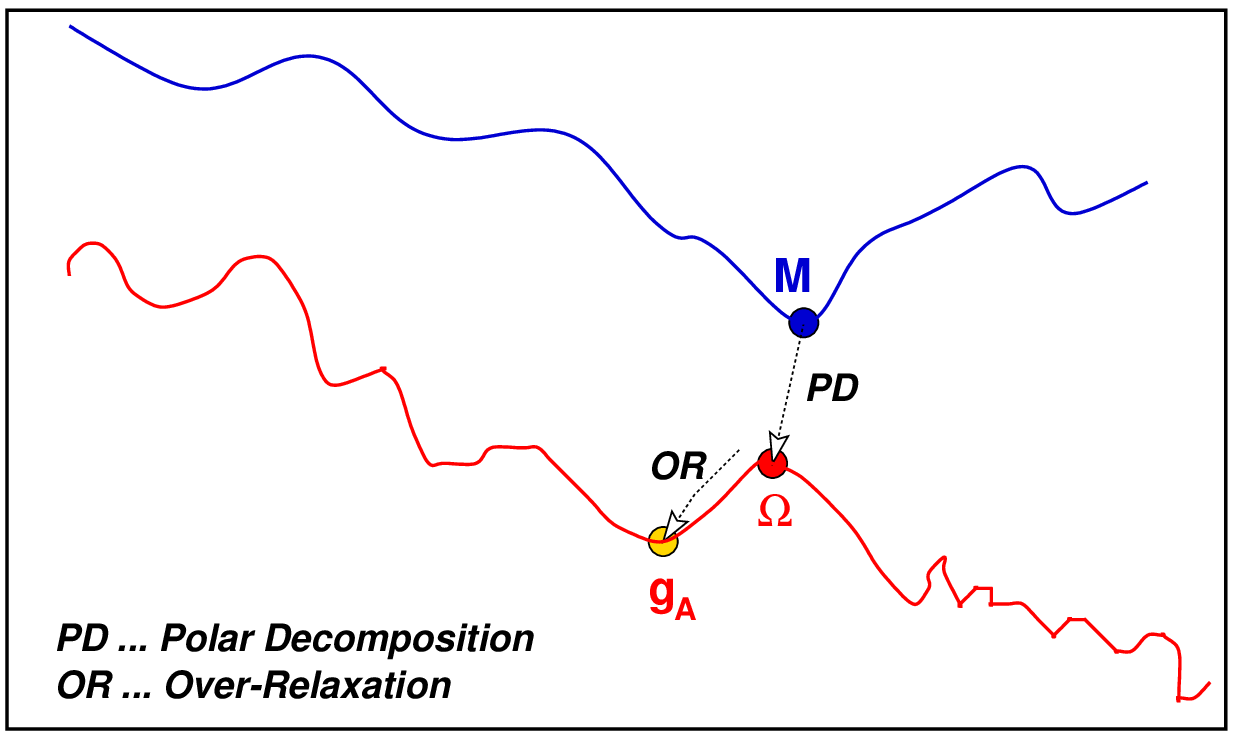}}}
\caption{Laplacian mapping of $M(x)$ to a nearby SO(3) matrix-valued
field $g_A(x)$.}
\label{scheme}
}
   It is interesting that a drawback of the over-relaxation technique,
in fixing to maximal center gauge, becomes a virtue in direct
Laplacian center gauge.  The drawback is that over-relaxation is not
good at finding the global minimum of $\S$, but only takes a given
initial state to a nearby local minimum (Gribov copy).  In fixing
to direct Laplacian gauge, on the other hand, we have taken advantage
of the fact that over-relaxation goes to a nearby minimum in order to
carry out the Laplacian mapping $M(x) \ra g(x)$.

 It may be useful at this stage to  summarize the steps of direct 
Laplacian center gauge fixing:
\begin{enumerate}
\item From a thermalized SU(2) lattice, construct the lattice
configuration in the adjoint representation
\beq
     [U_{A\m}(x)]_{ij} = \oh \mbox{Tr}[\s_i U_\m(x) \s_j U^\dg_\m(x)]
\eeq
\item Solve the eigenvalue problem \rf{eigen} for the eigenvectors
corresponding to the three smallest eigenvalues of the lattice
Laplacian operator in the adjoint representation.  
We have used the ARPACK routines \cite{ARPACK}
for this purpose.  From the eigenvectors $\vec{f^a}(x),~a=1,2,3$,
construct the matrix field $M_{ij}(x)=f_i^j(x)$.
\item Perform, at each site, the singular value decomposition
$M(x) = U(x) M_D(x) V^T(x)$, and extract the SO(3) matrix-valued field
\beq
   \Omega(x) = \det[U(x) V^T(x)] U(x) V^T(x)
\eeq
\item Map $\Omega$ to an SU(2) matrix-valued field
\beq
   \omega(x) = \pm {I + \s^a \s^b [\Omega(x)]_{ab} \over 2[1 + 
         \mbox{Tr}(\Omega(x))] }
\eeq
and perform the gauge transformation
\beq
    {}^\omega U_\m(x) = \omega^\dg(x) U_\m(x) \omega(x+\hat{\m})
\eeq
The result is to transform the configuration into Laplacian Landau
gauge in the adjoint representation.
\item With ${}^\omega U_\m(x)$ as the starting point, relax the
lattice field to the nearest (or, at least, a nearby) Gribov copy of direct
maximal center gauge, using the over-relaxation method described
in ref.\ \cite{Jan98}.  The result is the configuration fixed
to direct Laplacian center gauge.
\end{enumerate}

   To make a long story short, direct Laplacian center gauge in practice
is nothing but gauge fixing to adjoint Laplacian Landau gauge, followed
by over-relaxation to the nearest Gribov copy of direct maximal center
gauge.  Both the direct Laplacian and the direct maximal center gauges
aim at generating a configuration satisfying locally the adjoint lattice Landau
gauge condition.  The difference is that the optimal configuration, in
maximal center gauge, is generated by a gauge transformation maximizing
the $R$ functional, whereas in direct Laplacian center gauge 
the optimal configuration
is generated by a gauge transformation lying closest to a matrix maximizing
the $R_M$ functional.  We hope to have explained, in the previous section,
why the global maximum of $R$ is not necessarily the best choice for
vortex finding, and to have motivated the alternative choice
of Gribov copy corresponding to direct Laplacian center gauge.
In the next section, we present results that are obtained from center
projection in this gauge.

\section{Numerical Results}

   To test the reasoning of the last section, we have recalculated the
vortex observables introduced in our previous work (cf.\ ref.\
\cite{Us,Jan98}), with P-vortices located via center projection after
fixing the lattice to the new direct Laplacian center gauge.

\subsection{Center Dominance and Precocious Linearity}
   
   In view of the $\approx 30\%$ breakdown of center dominance in maximal
center gauge found by BKP \cite{BKP}, the quantities
of most immediate interest are the center-projected Creutz ratios,
extracted from $I\times J$ Wilson loops on the center-projected
lattice.  These are displayed, for $\b$ in the range $[0.4,2.5]$,
in Fig.\ \ref{nchi}.

\FIGURE[h!]{
\centerline{\scalebox{1.5}{\includegraphics{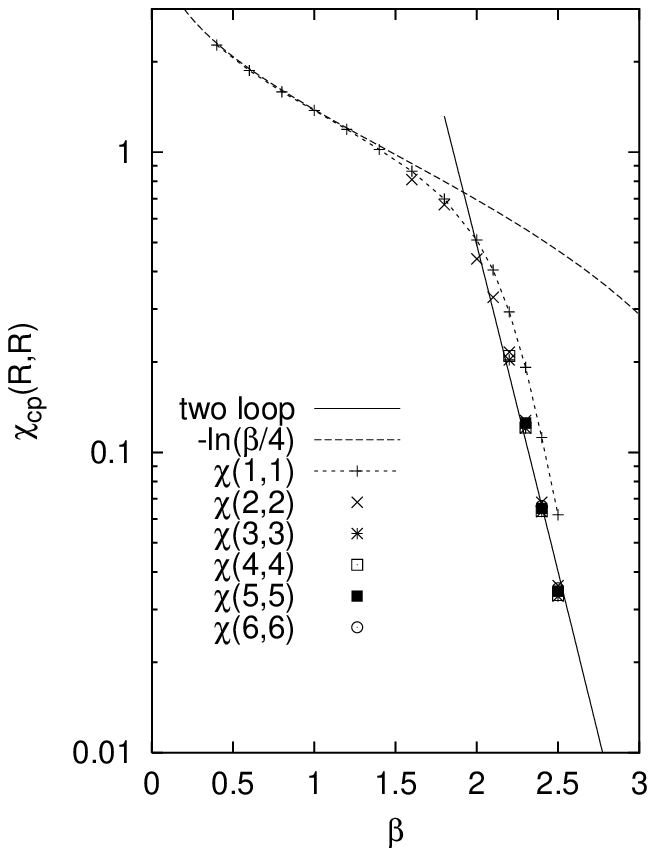}}}
\caption{Creutz ratios from center-projected lattice configurations,
in the direct Laplacian center gauge.}
\label{nchi}
}

    At strong couplings, it is clear from Fig.\ \ref{nchi} that
$\chi_{cp}(1,1)$ matches up with the analytic prediction, in the full
theory, of $\s = -\ln(\b/4)$.  At weaker couplings, we display our
data for $\b=2.4$ and $\b=2.5$ in Figs.\ \ref{2p4} and \ref{2p5}.
Once again we see the feature of precocious linearity; i.e.\ the fact
that Creutz ratios $\chi_{cp}(R,R)$ at a given $\b$ vary only slightly
with $R$, which means that the center-projected potential is
approximately linear starting at $R=2$ lattice spacings.  The
significance of precocious linearity is that it implies that the
center-projected degrees of freedom have isolated the long-range
physics, and are not mixed up with ultraviolet fluctuations.

\FIGURE[h!]{
\centerline{\scalebox{0.9}{\includegraphics{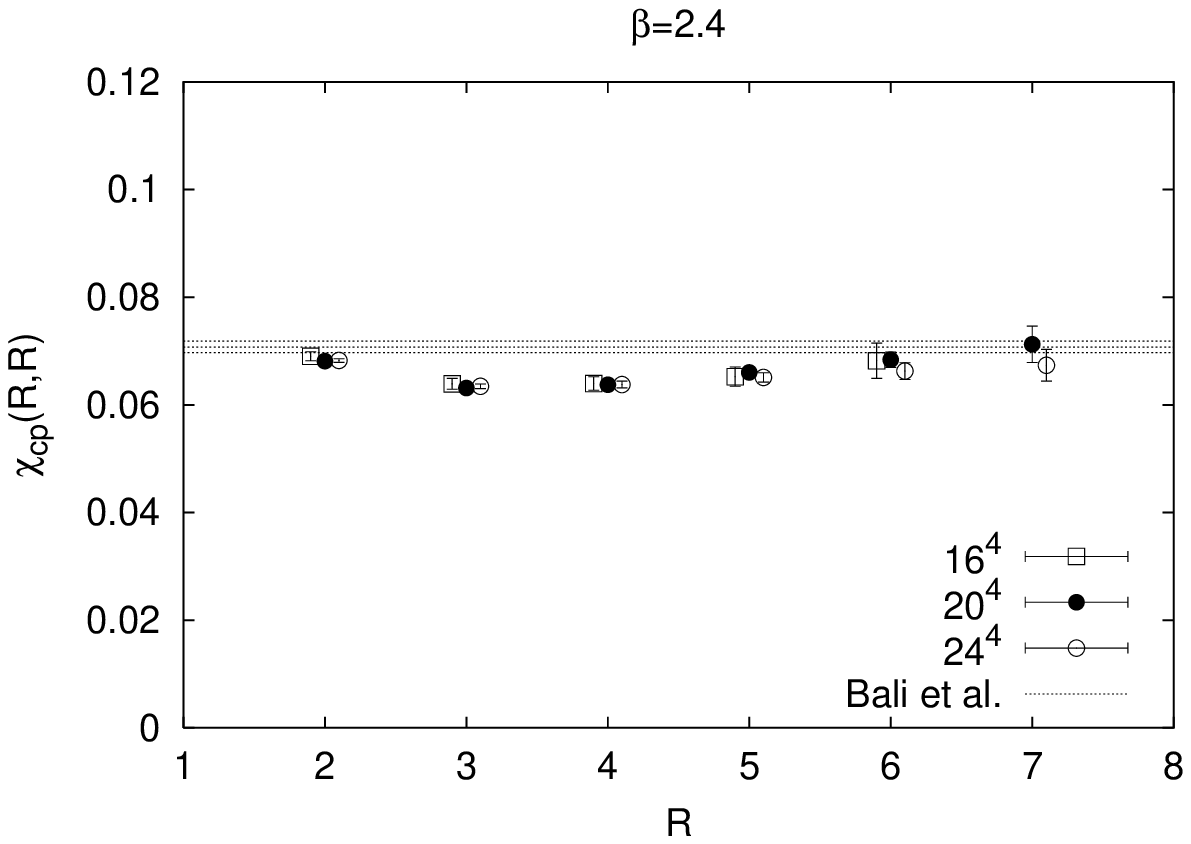}}}
\caption{Center projected Creutz ratios at $\b=2.4$ on various lattice
sizes. The horizontal
band indicates the full asymptotic string tension and errorbar.}
\label{2p4}
}

\FIGURE[h!]{
\centerline{\scalebox{0.9}{\includegraphics{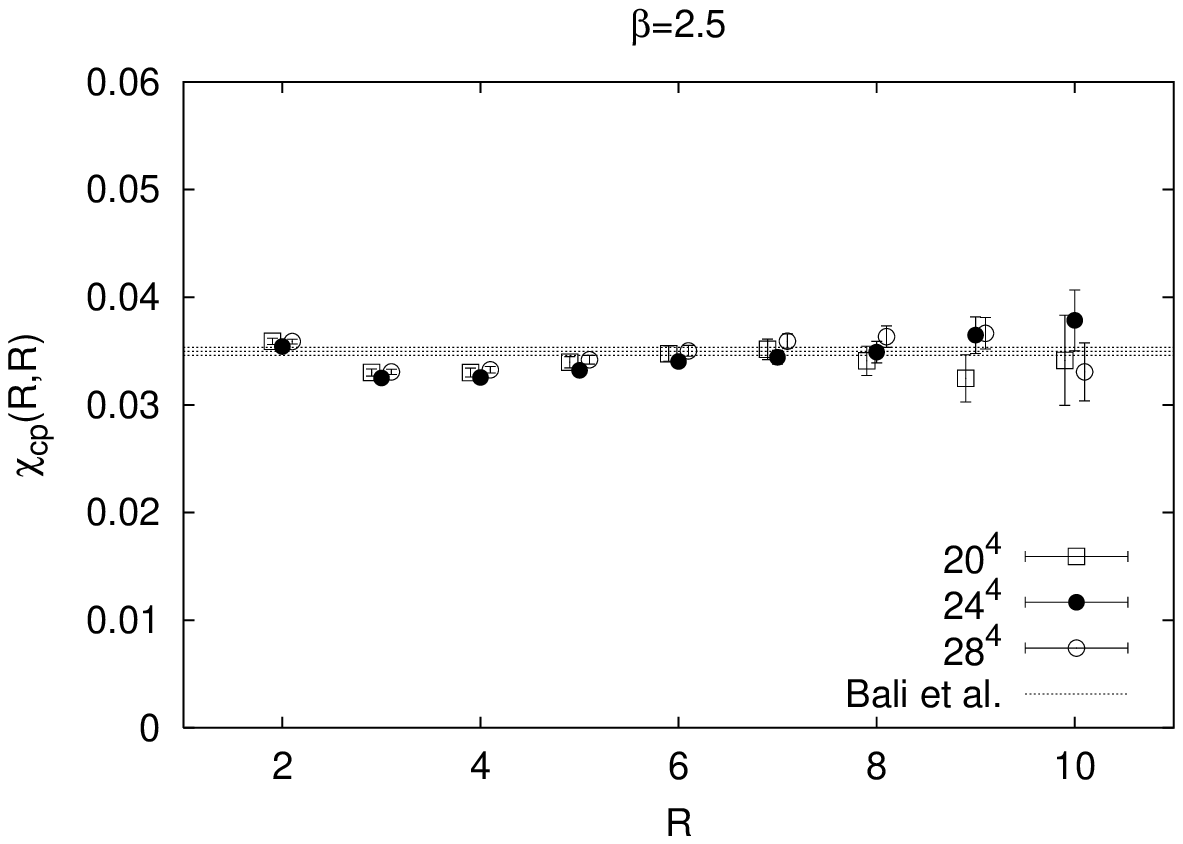}}}
\caption{Same as Fig.\ \ref{2p4}, at $\b=2.5$.}
\label{2p5}
}

   It is also apparent that the center-projected Creutz ratios are
quite close to the asymptotic string tension of the unprojected
theory, reported in refs. \cite{Bali}.  This is evidence, at least at
these two couplings, of the center dominance property.  Our combined
data for the range of couplings $\b=2.2-2.5$ is displayed on a
logarithmic plot in Fig.\ \ref{all_beta}.  In general
$\chi_{cp}(R,R)$ deviates from the full asymptotic string tension by
less than 10\%.

\FIGURE[h!]{
\centerline{\scalebox{1.1}{\includegraphics{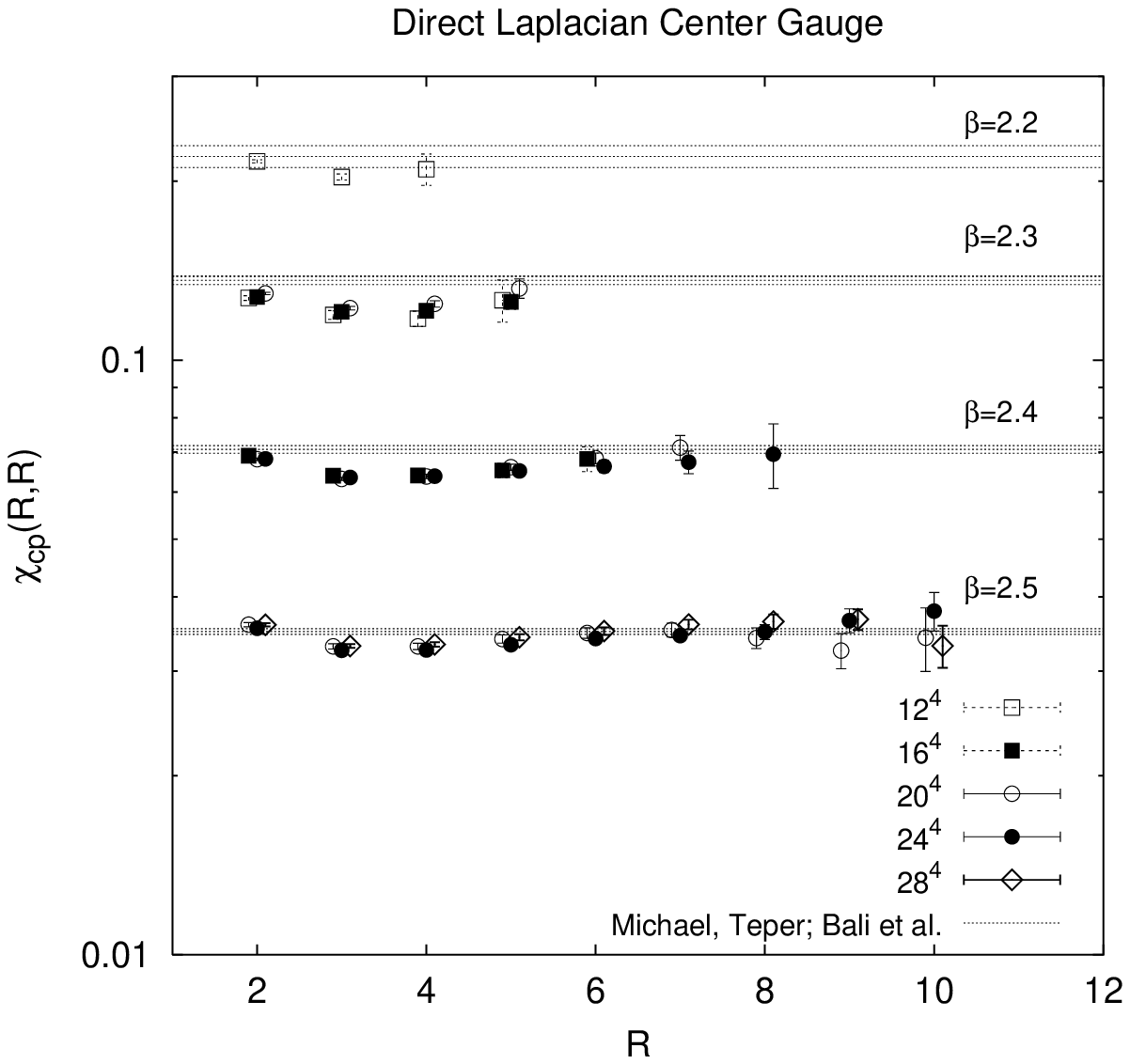}}}
\caption{Combined data, at $\b=2.2-2.5$, for center-projected
Creutz ratios obtained after direct Laplacian center gauge fixing.
Horizontal bands indicate the asymptotic string tensions on the
unprojected lattice, with the corresponding errorbars.}
\label{all_beta}
}

   As another way of displaying both center dominance and
precocious linearity, we adopt the usual procedure of assigning
a lattice spacing $a(\b)$ based on the asymptotic string tension
in lattice units $\s_{Lat}(\b)$, and the string tension in
physical units $\s_{phys}=(440~\mbox{MeV})^2$, i.e.
\beq
       a^2(\b) = {\s_{Lat}(\b) \over \s_{phys}}
\eeq
We then display, in Fig.\ \ref{sigma2}, the ratio 
\beq
   {\chi_{phys}(R,R) \over \s_{phys}} 
           = {\chi_{cp}(R,R) \over \s_{Lat}(\b)}
\eeq
as a function of the distance in physical units 
\beq
 R_{phys} = R a(\b)
\eeq 
for all $\chi_{cp}(R,R)$ data points
taken in the range of couplings $\b=2.3-2.5$.  Again we see that the
center-projected Creutz ratios and asymptotic string tension are in
good agreement (deviation $< 10\%$), and there is very little variation
in the Creutz ratios with distance.

\FIGURE[h!]{
\centerline{\scalebox{0.9}{\includegraphics{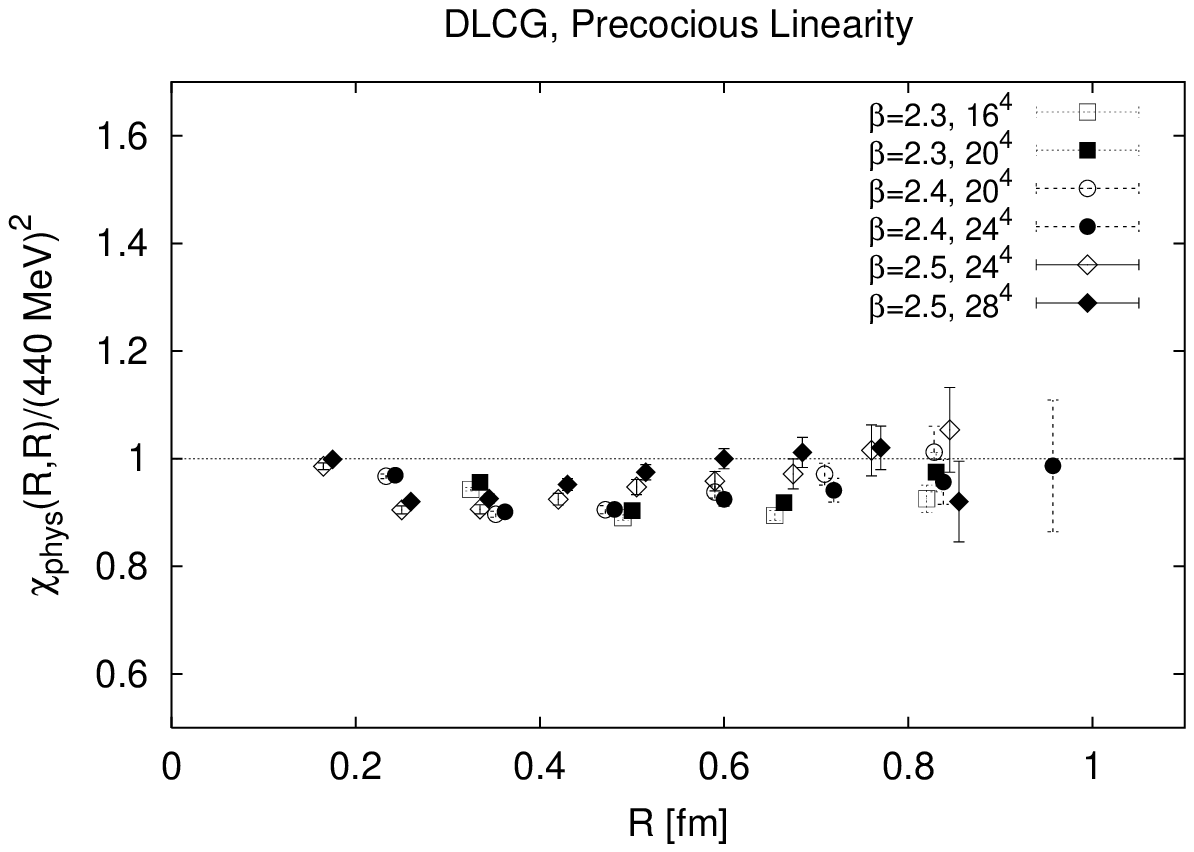}}}
\caption{The ratio of projected Creutz ratios to the full asymptotic
string tension, as a function of loop extension in fermis.  The
data is taken from $\chi_{cp}(R,R)$ at a variety of couplings and
lattice sizes.}
\label{sigma2}
}

   Precocious linearity can be understood in the following way:
Asymptotically, outside the vortex core, a center vortex is a $Z_N$
dislocation, which affects the value of large Wilson loops,
topologically linked to the vortex, by a center element factor.
Assume that P-vortices correctly identify the middle of thick center
vortices in lattice configurations.  Then center projection in effect
collapses the thick ($\approx$ one fermi) core of a vortex to a width
of one lattice spacing.  This means that the asymptotic effect of
thick vortices is obtained, on the projected lattice, at any distance
greater than one lattice spacing.  If P-plaquettes in a plane are
uncorrelated, this leads to a linear potential at short distances on
the projected lattice.  On the other hand, if precocious linearity is
not found on the projected lattice, it means that either the P-vortex
surface is very rough, fluctuating on all distance scales, or else
that some large fraction of P-plaquettes on the projected lattice
belong to P-vortices which are small in extent, and do not percolate.
Either case results in some short-range correlations among P-plaquettes in
a plane, corresponding to high-frequency phenomena not directly
associated with the long-range physics.

\subsection{Scaling of the Vortex Density}
   
   Denote the total number of plaquettes on the lattice
by $N_T$, the number of P-plaquettes by $N_{vor}$, and the
density $N_{vor}/N_T$ of P-plaquettes on the lattice by $p$.
We have
\bea
      p       &=&
                  {N_{vor}\over N_T} = {N_{vor}a^2 \over N_T a^4} a^2
\non \\
              &=& 
              {\mbox{Total Vortex Area} \over 6
                  \times \mbox{Total Volume} } a^2
\non \\
              &=& {1\over 6} \rho a^2
\non
\eea
where $\rho$ is the center vortex density in physical units. Then,
according to asymptotic freedom
\bigskip

\beq
     p = {\rho \over 6\Lambda^2} \left({6\pi^2 \over 11} \b \right)^{102/121}
            \exp\left[-{6\pi^2 \over 11}\b \right]
\label{afree}
\eeq
The average vortex density $p$ can be extracted from the expectation value
of center projected plaquettes
\bea
     W_{cp}(1,1) &=& (1-p)\times (+1) + p \times (-1) = 1-2p
\non \\ 
        p &=& \oh ( 1 - W_{cp}(1,1) ) 
\eea

   Figure \ref{pvor} is a logarithmic plot of P-vortex density $p$
vs.\ $\b$.  Errorbars are less than the size of data points.
The solid line is the asymptotic freedom prediction   
\rf{afree}, with $\sqrt{\rho / 6\Lambda^2} = 50$.  We emphasize that
the slope of this line represents the proper asymptotic
scaling for surface densitites.
The slope that would be associated with the scaling of pointlike
objects such as instantons, or linelike objects such as monopoles,
would be quite different. Note that the apparent scaling of 
$\chi_{cp}(1,1)=-\ln W_{cp}(1,1)\approx 2p$ in Fig.\ \ref{nchi} is
just a consequence of the scaling of $p$. 

\FIGURE[h!]{
\centerline{\scalebox{1.5}{\includegraphics{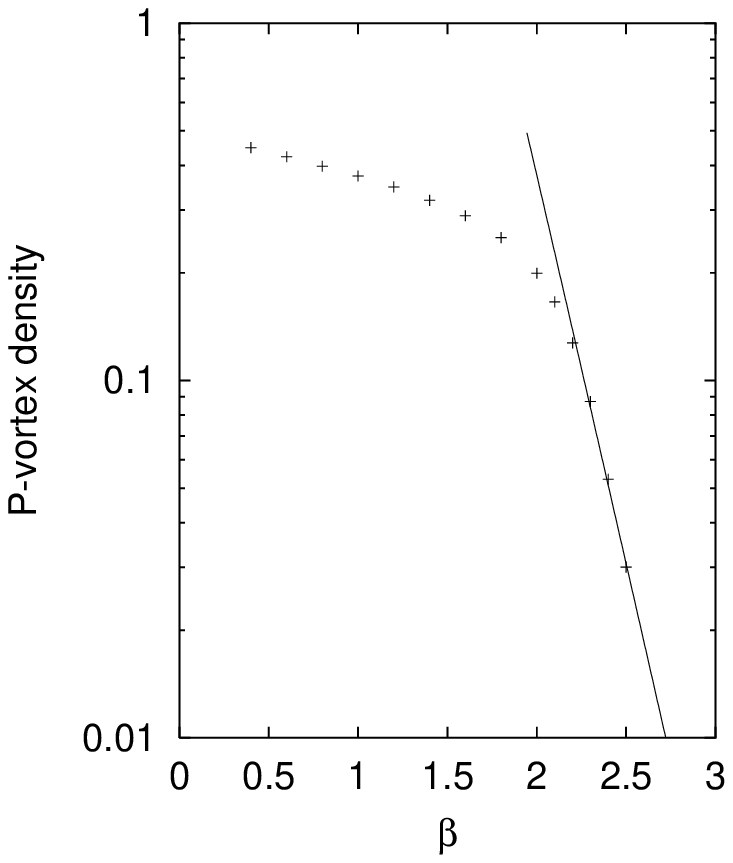}}}
\caption{Evidence of asymptotic scaling of the P-vortex surface
density.  The solid line is the asymptotic freedom prediction 
\rf{afree}, with $\sqrt{\rho / 6\Lambda^2} = 50$.}
\label{pvor}
}

\subsection{Vortex-Limited Wilson Loops}

   A ``vortex-limited'' Wilson loop $W_n(C)$ is defined as the
expectation value of an unprojected Wilson loop around some 
contour $C$, evaluated in the sub-ensemble of configurations 
in which, on the corresponding center-projected lattice, precisely
$n$ P-vortices pierce the minimal area of the loop.  If P-vortices
on the projected lattice roughly locate the middle of thick center
vortices on the unprojected lattice, then in the limit of large
loop areas, for the SU(2) gauge group, we expect \cite{Jan98,Us}
\beq
     {W_n(C) \over W_0(C)} \longrightarrow (-1)^n
\eeq
The numerical evidence definitely shows a trend in this direction,
as can be seen from our data for $W_1/W_0$ and $W_2/W_0$ vs.\ minimal
loop area at $\b=2.3$, shown in Figs.\ \ref{vtex1} and \ref{vtex2}.

\FIGURE[h!]{
\centerline{\scalebox{0.65}{\includegraphics{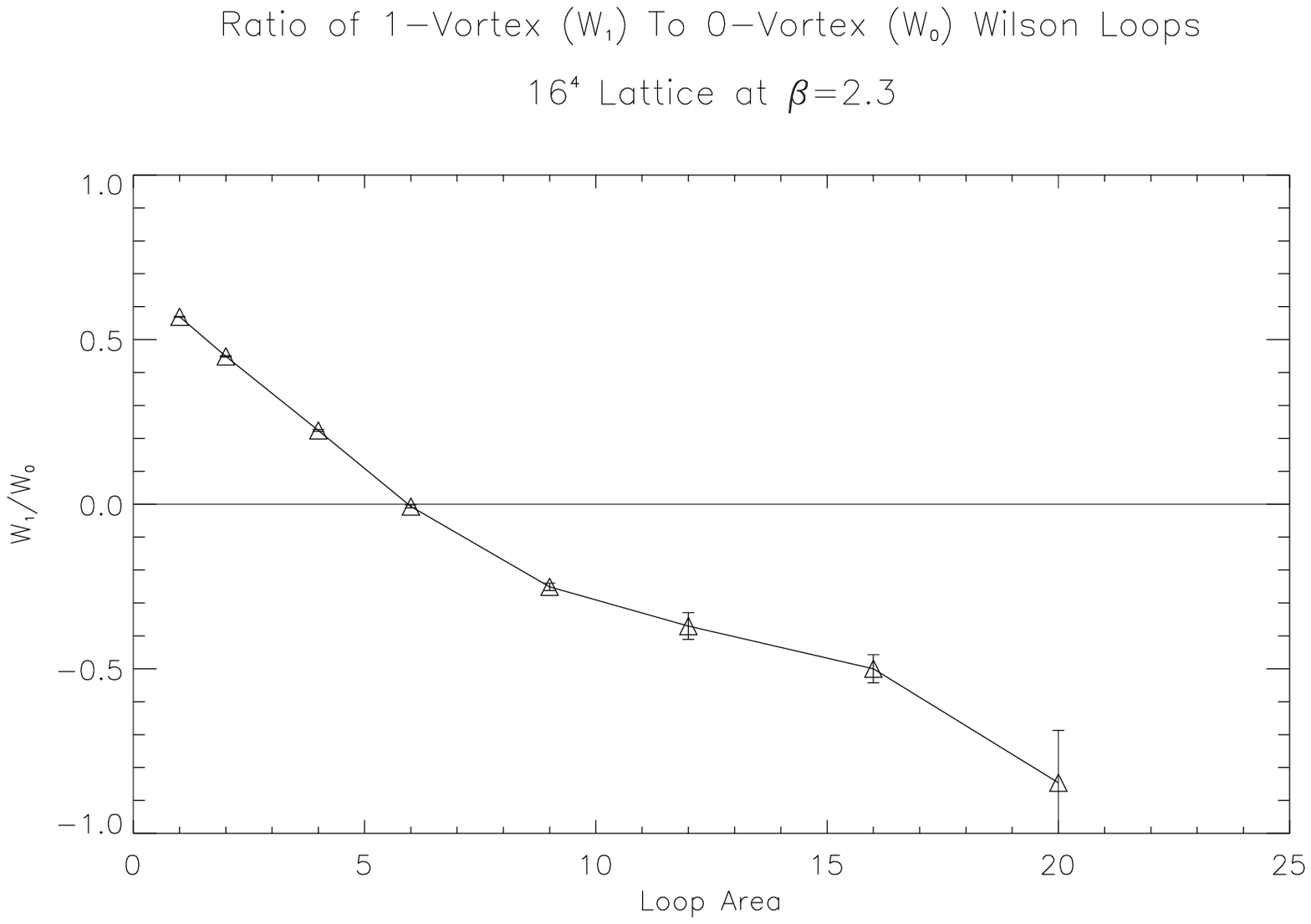}}}
\caption{Ratio of the one-vortex to zero-vortex Wilson loops
$W_1(C)/W_0(C)$ vs.\ loop area, at $\b=2.3$ on a $16^4$ lattice.}
\label{vtex1}
}

\FIGURE[h!]{
\centerline{\scalebox{0.65}{\includegraphics{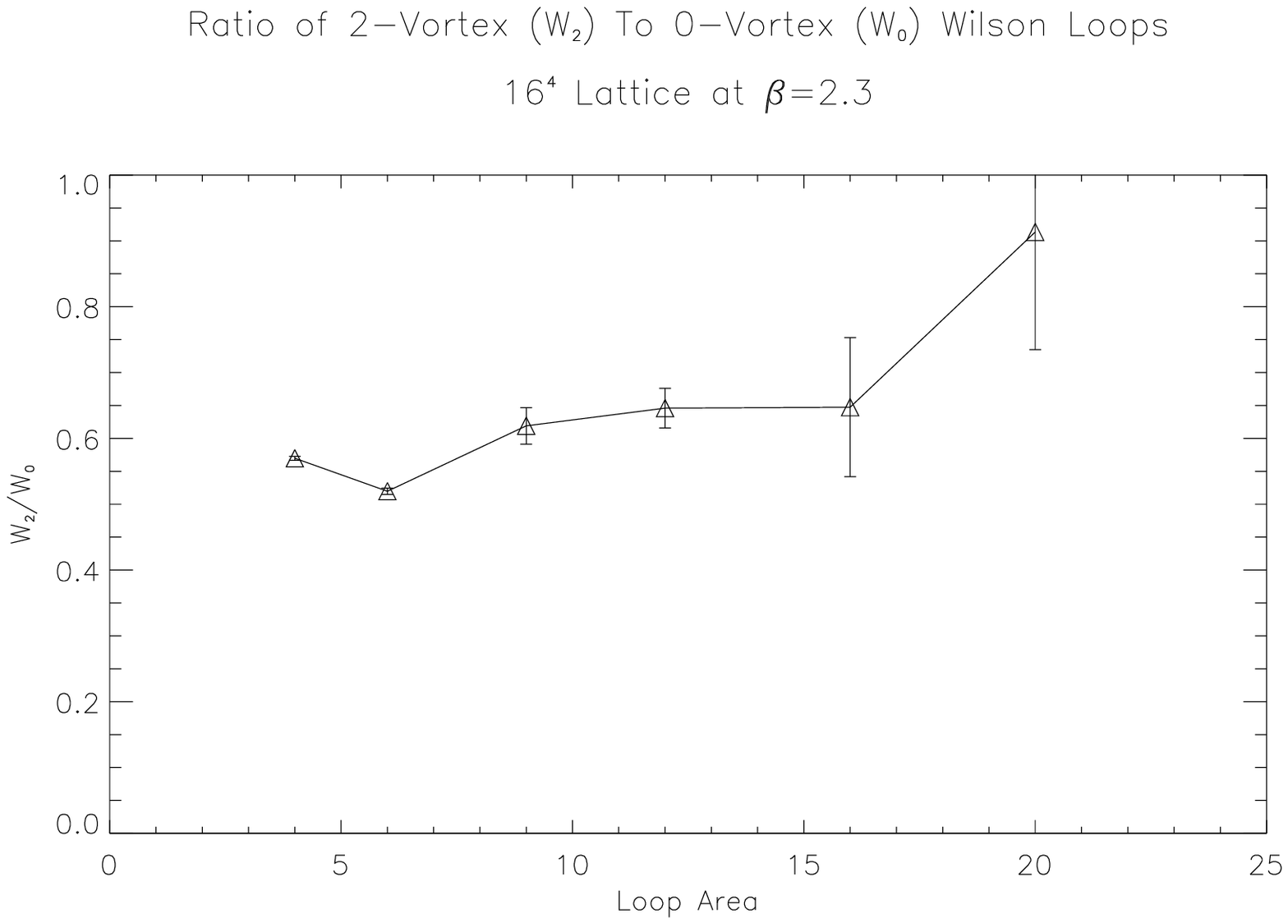}}}
\caption{Ratio of the two-vortex to zero-vortex Wilson loops
$W_2(C)/W_0(C)$ vs.\ loop area, at $\b=2.3$ on a $16^4$ lattice.}
\label{vtex2}
}

   If P-vortices in the projected configuration locate thick
center vortices in the unprojected configurations, and if
center vortices are responsible for confinement, then $W_0(C)$
should not have an area-law falloff.  In Fig.\ \ref{nchi0}
we compare Creutz ratios $\chi_0[I,J]$
extracted from rectangular $W_0(I,J)$ loops, with the standard
Creutz ratios $\chi(I,J)$ from loops evaluated in the full
ensemble.  As expected, the Creutz ratios of the zero-vortex
loops tend to zero with loop area.

\FIGURE[h!]{
\centerline{\scalebox{0.7}{\includegraphics{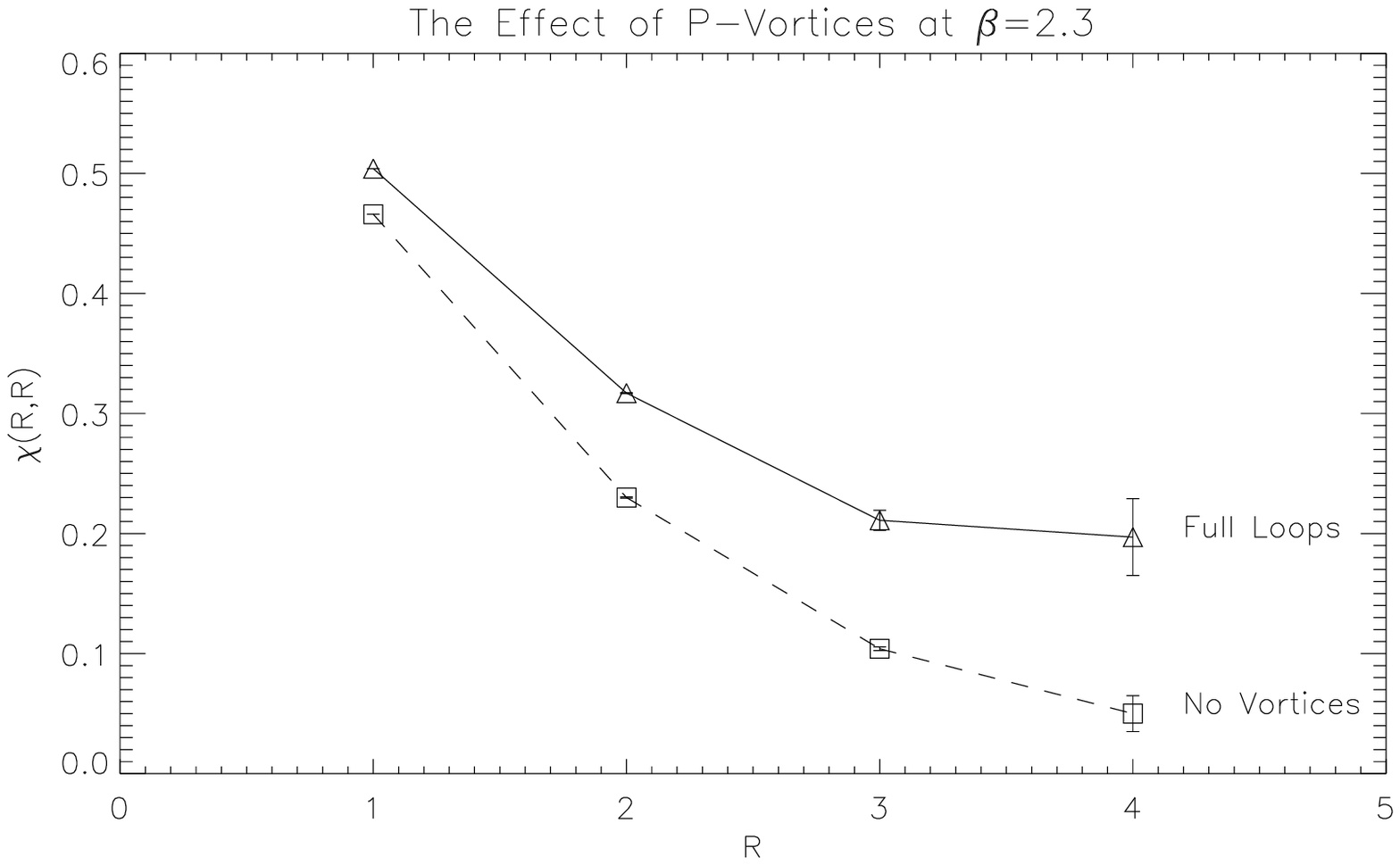}}}
\caption{Creutz ratios $\chi_0(R,R)$ extracted from zero-vortex
Wilson loops $W_0(I,J)$, as compared to the usual Creutz ratios
$\chi(R,R)$ on a $16^4$ lattice at $\b=2.3$.}
\label{nchi0}
}

\subsection{Center Vortex Removal}

   It was suggested by de Forcrand and D'Elia \cite{dF1}
that one could remove center vortices from a given lattice
configuration by simply multiplying that configuration
by the corresponding center-projected
configuration derived in maximal center gauge, i.e.\
\beq
       U'_\m(x) \equiv Z_\m(x) U_\m(x)
\label{remove}
\eeq
where $Z_\m(x)$ is given by \rf{Z}.  Since the adjoint representation
is blind to center elements, it is easy to see that if $g(x)$ is a 
transformation such that ${}^gU_\m(x)$ is in maximal center gauge, 
then ${}^gU'_\m(x)$ is 
also in maximal center gauge.  However, there are no P-vortices
obtained from center projection of ${}^gU'_\m$, since
\beq 
    Z'_\m(x)=\mbox{signTr}[{}^gU'_\m(x)]=  Z^2_\m(x) = 1
\eeq
One can therefore say that center vortices, as identified
in maximal center gauge, have been removed from the
lattice configuration.  More precisely, what the modification
\rf{remove} does is to place a thin vortex (one plaquette thickness)
in the middle of
each thick center vortex core, whose locations are identified by
center projection.  At large scales, the effects of the thin
and thick vortices on Wilson loops will cancel out.
Thus there should be no area law due
to vortices in the modified configuration $U'_\m$, and the
asymptotic string tension should vanish.  

   The vanishing of string tension in the modified configurations
was, in fact, observed in ref.\ \cite{dF1}, using maximal center gauge fixing
by the over-relaxation technique of ref.\ \cite{Jan98}.
Figure \ref{forc} is a repeat of the de Forcrand-D'Elia 
calculation at $\b=2.3$, using
direct Laplacian center gauge rather than direct maximal center
gauge, and we find essentially the same result.

\FIGURE[h!]{
\centerline{\scalebox{0.9}{\includegraphics{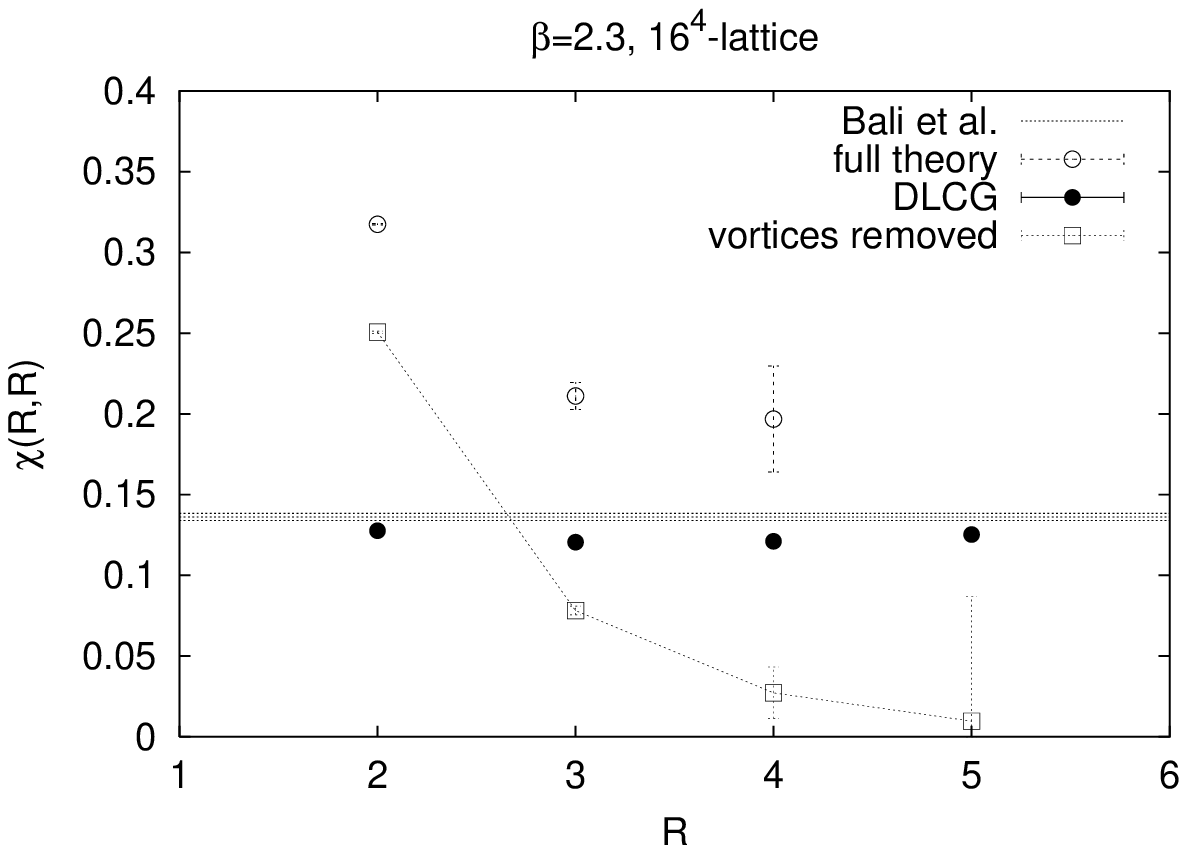}}}
\caption{Creutz ratios on on the modified
lattice, with vortices removed, at $\b=2.3$.  
For comparison, we also display the unprojected Creutz ratios
(open circles), the center projected Creutz ratios (solid circles), and
the asymptotic string tension (horizontal band).}
\label{forc}
}


\section{Remarks on Gauge-Fixing Ambiguities}

   The original suggestion of 't Hooft, in his 1981 article on
monopole confinement \cite{thooft}, 
was to introduce a composite operator in the
pure gauge theory transforming like a field in the adjoint representation
of the gauge group; for SU(2) gauge theory this operator 
defines a unitary gauge leaving a residual U(1) symmetry. 
Monopole worldlines would then be associated with
lines along which the gauge transformation is ambiguous.  This was the idea
which motivated the numerical study of maximal abelian \cite{mag}
and Laplacian abelian \cite{lag} gauges. The idea can be
generalized by introducing not one but two composite operators transforming
in the adjoint representation.  Let us denote these operators, in
SU(2) gauge theory, by $\phi_a(x)$ and $\eta_a(x)$, with color
index $a=1,2,3$.  One then performs a gauge transformation which takes,
e.g., $\phi_a$ into the positive color 3 direction at every point, and 
$\eta_a$ to lie in the color 1-3 plane, with $\phi_3,\eta_1 > 0$.  
This gauge leaves a remnant $Z_2$ symmetry.  Gauge-fixing ambiguities
occur on surfaces where $\phi(x)$ and $\eta(x)$ are co-linear, and
these surfaces are then to be identified with center vortices. 

   The suggestion of refs.\ \cite{Alex,Pepe} is to choose, for $\phi_a(x)$ and
$\eta_a(x)$, the two lowest lying eigenstates of the covariant
lattice Laplacian operator \rf{D} in the adjoint representation; i.e.
\beq
     \phi_a(x) \equiv f^1_a(x) ~~,~~~  \eta_a(x) \equiv f^2_a(x) 
\eeq
with $\l_1<\l_2$ being the two lowest eigenvalues of the Laplacian
operator.  This is the original version of Laplacian center gauge, which
we will refer to as LCG1.  Because the procedure involves first fixing
to Laplacian abelian gauge, followed by a further gauge fixing which
reduces the residual gauge symmetry from U(1) to $Z_2$, LCG1 is
reminiscent of the indirect maximal center gauge of ref.\ \cite{Us}.
As we have pointed out in refs.\ \cite{zako}, abelian monopole
worldlines in the indirect maximal center 
gauge lie on center vortex surfaces, and
a vortex at fixed time can be viewed as a monopole-antimonopole chain.  
The same relationship between abelian monopoles and center vortices holds
true in LCG1.

  Gauge-fixing ambiguities in LCG1 occur when $\phi$ and $\eta$ are
co-linear in color space, and it is suggested that these ambiguities
can be used to locate vortex surfaces.  This approach to
vortex finding is certainly quite different
from the reasoning outlined in section 3, which is is motivated by a
``best fit'' procedure.  In special cases, notably for a classical vortex
solution on an asymmetric lattice with twisted boundary conditions, the
co-linearity approach seems to work well \cite{Montero}.  We take note,
however, of a simple counter-example: Suppose we 
insert two vortex sheets ``by
hand'' into a configuration $U_\m(x)$ by the transformation
\beq
        U_0(x) \ra U'_0(x) = \left\{ \begin{array} {cl}
              - U_0(x) & ~~ x_0=0 ~\mbox{and}~ 0<x_1<L \cr
                U_0(x)  & ~~ \mbox{otherwise} \end{array} \right.
\eeq
All other components are unchanged, i.e.\ $U'_k(x)=U_k(x)$ for $k>0$.  This
transformation inserts two thin vortex sheets into the lattice, parallel
to the $x_2-x_3$ plane, at $x_0=0$.  We then ask whether these
two vortex sheets will be located by the gauge-fixing ambiguity approach.
The answer is clearly no, since the composite Higgs fields 
$\phi(x)$ and $\eta(x)$ are identical in the original $U_\m(x)$, and in the
modified $U'_\m(x)$ configurations.  This example illustrates the fact that
the gauge-ambiguity approach lacks the ``vortex-finding property''
discussed in ref.\ \cite{vf},
which is the ability of a procedure
to locate thin vortices inserted at known locations
into the lattice.  We have argued in ref.\ \cite{vf} that this property
should be a necessary (although not sufficient) condition for locating 
vortices on thermalized lattices.

   In practice, on thermalized lattices, vortices are located in LCG1
via center projection, rather than by eigenvector co-linearity
\cite{Pepe}.  Using center projection to locate vortices, LCG1
recovers the vortex-finding property, for reasons explained in ref.\
\cite{vf}.\footnote{As discussed in that reference, the vortex-finding
property is obtained from center projection in any gauge which
completely fixes link variables in the adjoint representation, leaving
a residual $Z_N$ invariance.}  It also exhibits center dominance of the
projected asymptotic string tension.  On the other hand, 
the projected Creutz ratios in
LCG1 do not display precocious linearity \cite{Pepe}, nor does the
vortex surface density scale according to the asymptotic freedom
formula \cite{Tubby}.  A possible remedy is to first fix the lattice
to LCG1, and from there to fix to (direct or indirect) maximal center
gauge by an over-relaxation procedure.  This latter procedure has, in
fact, been tried by Langfeld et al.\ in ref.\ \cite{Tubby}, with good
results for the vortex density.  Other vortex observables, discussed
in the previous section, have not yet been studied systematically in
this approach, which seems to have a great deal in common with the direct
Laplacian center gauge we have advocated here.

\section{Conclusions}

   We have tested a procedure for locating center vortices on
thermalized lattices, based on the idea of finding the best fit to the
thermalized lattice by thin vortex configurations.  Our new procedure,
which is essentially just a variation of direct maximal center gauge,
is designed to soften the inevitable 
bad fit at vortex cores, due to the singular
field strength of thin vortices.  The numerical results we have found
are promising: Deviations from center dominance are generally less
than 10\%, P-vortex density scales correctly, and there are the usual
strong correlations between P-vortex locations and gauge-invariant
observables.  Our new method is motivated by an improved understanding of
how maximal center gauge works, and it addresses some objections to center
gauge fixing that have been raised in the recent literature.  We
hope it will provide a more solid foundation for further numerical
studies.

\acknowledgments{ Our research is supported in part by
Fonds zur F\"orderung der
Wissenschaftlichen Forschung P13997-PHY (M.F.), the U.S. Department of 
Energy under Grant No.\ DE-FG03-92ER40711 (J.G.), and the Slovak Grant 
Agency for Science, Grant No. 2/7119/2000 (\v{S}.O.).
Our collaborative effort is also supported by NATO Collaborative Linkage
Grant No.\ PST.CLG.976987. }

\end{document}